\documentclass[showpacs,aps,prd,preprint,nofootinbib,showkeys,unsortedaddress,raggedbottom]{revtex4-1}
\pdfoutput=1

\usepackage{bm}
\usepackage{amsmath}
\usepackage{graphicx}
\usepackage{subfigure}
\usepackage[usenames,dvipsnames]{color}
\definecolor{darkblue}{RGB}{0,0,196}
\usepackage[colorlinks=true,linkcolor=darkblue,citecolor=darkblue,urlcolor=darkblue]{hyperref}

\usepackage{setspace}
\usepackage{footmisc}
\usepackage[makeroom]{cancel}

\renewcommand{\vec}[1]{\mathbf{#1}}

\def\be{\begin{equation}}
\def\ee{\end{equation}}
\def\ba{\begin{eqnarray}}
\def\ea{\end{eqnarray}}

\begin{document}

\title{Pion interferometry at 200 GeV using anisotropic hydrodynamics}

\author{Mubarak Alqahtani} 

\affiliation{Department of Basic Sciences, College of Education, Imam Abdulrahman Bin Faisal University, Dammam 34212, Saudi Arabia}

\author{Michael Strickland}

\affiliation{Department of Physics, Kent State University, Kent, OH 44242 United States}

\begin{abstract}

In this paper, we continue our phenomenological studies of heavy-ion collisions using 3+1d anisotropic hydrodynamics (aHydro). In previous works, we compared quasiparticle aHydro (aHydroQP) with ALICE 2.76 TeV Pb-Pb and RHIC 200 GeV Au-Au collision results. At both energies, the agreement  was quite good between aHydroQP and the experimental data for many observables. In this work, we present comparisons of the Hanbury Brown--Twiss (HBT) radii and their ratios determined using $\pi^+ \pi^+$ pairs produced in 200 GeV Au-Au collisions. We first present comparisons with STAR results for the HBT radii and their ratios. We then present comparisons with PHENIX results for the HBT radii and their ratios. In both cases, we find reasonable agreement between aHydroQP predictions and available experimental results for the ratios of HBT radii.  At the level of the radii themselves, in some cases quantitative differences on the order of 10-20\% remain, which deserve further study.
\end{abstract}

\date{\today}



\maketitle

\section{Introduction}
\label{sec:intro}

Heavy-ion collision experiments at the Relativistic Heavy Ion Collider (RHIC) and Large Hadron Collider (LHC) create and study the quark-gluon plasma (QGP) under very extreme conditions. Relativistic dissipative hydrodynamics has been quite successfully used to describe the collective behavior seen in these experiments \cite{Huovinen:2001cy,Romatschke:2007mq,Ryu:2015vwa,Niemi:2011ix,Averbeck:2015jja,Jeon:2016uym,Romatschke:2017ejr}. A decade ago, anisotropic hydrodynamics was introduced as a new approach which takes into account the fact that the QGP is a highly momentum-space
anisotropic plasma at early times and near its longitudinal/transverse edges~\cite{Florkowski:2010cf,Martinez:2010sc,Martinez:2012tu,Ryblewski:2012rr,Alqahtani:2017mhy}. It was only recently that anisotropic hydrodynamics was extended to include both shear and bulk viscous effects and a 3+1d code was developed which enabled practioners to perform phenomenological comparisons~\cite{Alqahtani:2017jwl, Alqahtani:2017tnq, Almaalol:2018gjh}.

In recent years, using our quasiparticle anisotropic hydrodynamics code (aHydroQP), we presented phenomenological comparisons with data available at different collision energies. We first showed comparisons with 2.76 TeV Pb-Pb collision data from the ALICE collaboration \cite{Alqahtani:2017jwl, Alqahtani:2017tnq}. We presented comparisons of charged-hadron multiplicity, identified-particle spectra, identified-particle average transverse momentum, charged-particle elliptic flow, identified-particle elliptic flow, the integrated elliptic flow vs pseudorapidity, and the HBT radii. These comparisons showed that the agreement is quite good between our model and the experimental data.   Subsequently, we presented comparisons with 200 GeV Au-Au collision data from the RHIC experiments \cite{Almaalol:2018gjh}. In this prior work, we presented comparisons of the identified particle spectra, charged particle multiplicity versus pseudorapidity, identified particle multiplicity versus centrality, identified particle elliptic flow versus transverse momentum, and charged particle elliptic flow as a function of transverse momentum and rapidity.

In this paper, we will present aHydroQP predictions for Hanbury Brown-Twiss (HBT) radii in 200 GeV Au-Au collisions.  HBT interferometry studies the correlations of pairs of particles obeying Bose-Einstein statistics and is based on quantum statistical interference. The understanding of these correlations is crucial to study the system's dynamics and the space-time structure of the emitting sources formed at freeze-out \cite{Wiedemann:1999qn, Heinz:1999rw, Lisa:2005dd}.   We present comparisons of the HBT radii and their ratios for $\pi^+ \pi^+ $ pairs in different centrality classes as a function of $k_T$ (pair mean transverse momentum). The dependence  of the HBT radii on $k_T$ provides information about the size of the regions over which the system emits particles with similar momentum (region of homogeneity).  We first present comparisons with STAR results for the HBT radii and their ratios.   We then show comparisons with PHENIX results from 2015 for the HBT radii and their ratios. In both cases, we find reasonable agreement between our model and the experimental results, however, quantitative differences remain.  One should note that there are prior works which were able to describe results from pion interferometry at 200 GeV Au-Au collisions ~\cite{Broniowski:2008vp,Kisiel:2008ws,Pratt:2008sz,Pratt:2008qv,Bozek:2010er,Karpenko:2012yf,Bozek:2014hwa,Moreland:2015dvc}.  In this paper, we want to investigate the effect of the aHydro anisotropic distribution function on the HBT radii using our previously obtained parameter set, since the HBT correlations are sensitive to the freeze-out conditions.

The structure of the paper is as follows.  In Sec.~\ref{sec:aHydroQP}, we review the basics of anisotropic hydrodynamics and the 3+1d dynamical equations. In Sec.~\ref{sec:3+1daHydroQP}, we summarize the main components of the 3+1d aHydroQP model, which was used previously and will be used in this work. In Sec.~\ref{sec:results}, for Au-Au collisions at 200 GeV, we show comparisons of the HBT radii and their ratios obtained using the 3+1d aHydroQP model with experimental data. Sec.~\ref{sec:conclusions} contains our conclusions and an outlook for the future. 

\section{Quasiparticle Anisotropic Hydrodynamics}
\label{sec:aHydroQP}

The evolution of massive relativistic quasiparticle systems is governed by the Boltzmann equation~\cite{Alqahtani:2015qja}

\be
p^\mu \partial_\mu f(x,p)+\frac{1}{2}\partial_i m^2\partial^i_{(p)} f(x,p)=-C[f(x,p)]\,,
\label{eq:boltzmanneq}
\ee
where the mass ($m$) is a function of temperature obtained from lattice QCD calculations and $C[f(x,p)]$ is the collisional kernel which is taken to be in relaxation-time approximation (RTA) \cite{Alqahtani:2015qja}. 

In anisotropic hydrodynamics,  the distribution function in the local rest frame is given by 
\be
f(x,p) =  f_{\rm eq}\!\left(\frac{1}{\lambda}\sqrt{\sum_i \frac{p_i^2}{\alpha_i^2} + m^2}\right) ,
\label{eq:fform}
\ee
where the  $\alpha_i$ parameters encode the momentum-space anisotropy of the medium.  This form reduces back to the equilibrium distribution function with temperature $T$ when $\alpha_i=1$ and the temperature like parameter, $\lambda$,  is taken to be $\lambda$ = T.  Once the form of the distribution function is specified, the dynamical equations can be obtained using the first and second moments of the Boltzmann equation. Detailed derivation of the dynamical equations for aHydroQP can be found in Refs.~\cite{Nopoush:2014pfa,Alqahtani:2015qja,Alqahtani:2016rth,Alqahtani:2017mhy}.

\begin{figure}[t!]
\centerline{
\hspace{-1.5mm}
\includegraphics[width=1\linewidth]{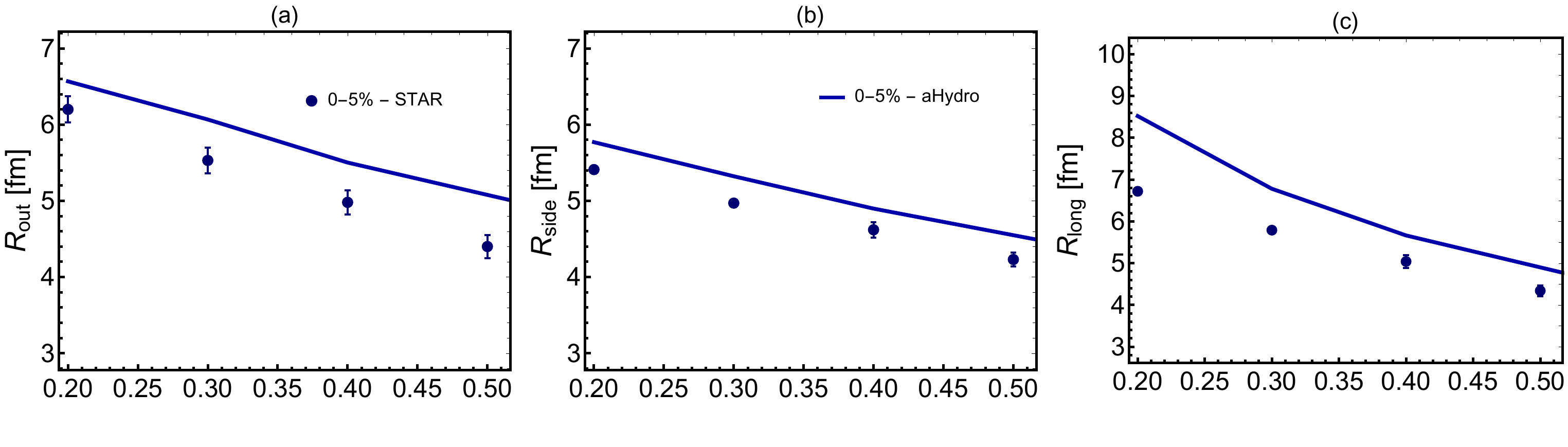}}
\centerline{
\hspace{-1.5mm}
\includegraphics[width=1\linewidth]{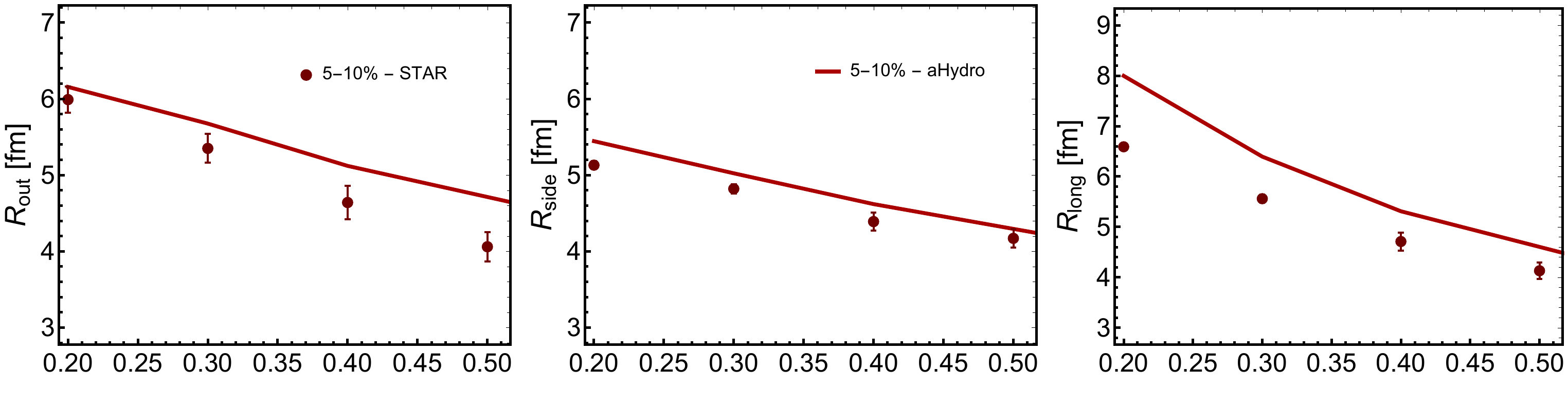}}
\centerline{
\hspace{-1.5mm}
\includegraphics[width=1\linewidth]{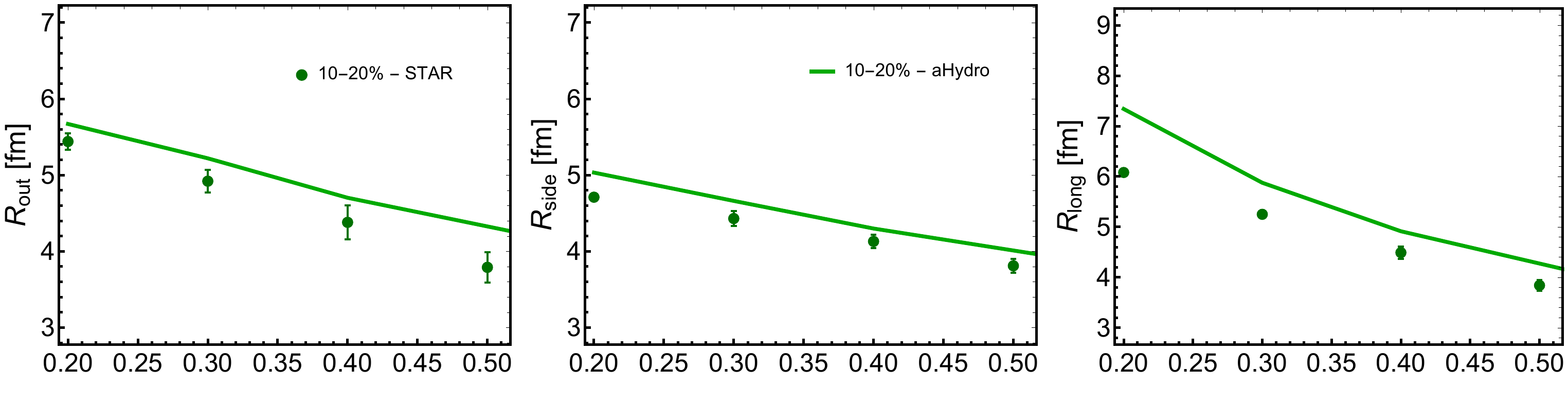}}
\centerline{
\hspace{-1.5mm}
\includegraphics[width=1\linewidth]{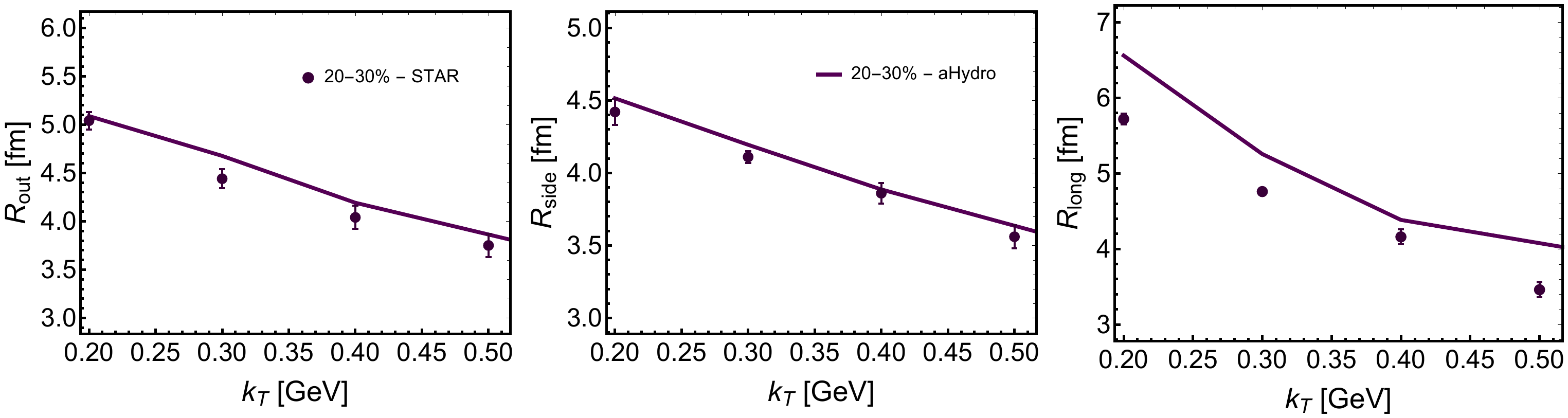}}
\caption{HBT radii are as a function of  $k_T$ for $\pi^+ \pi^+ $ in the  0-5$\%$, 5-10$\%$, 10-20$\%$, and 20-30$\%$ centrality classes. The left, middle, and right columns show $R_{\rm out} $, $ R_{\rm side} $, and $R_{\rm long}$, respectively. All results are for 200 GeV Au-Au collisions and data shown are from the STAR collaboration \cite{Adams:2004yc}. The $p_T$ cuts used were $0.15 < p_T < 0.8$~GeV~\cite{Adams:2004yc}. }
\label{fig:HBTSTAR}
\end{figure}

\section{3+1d aHydroQP model}
\label{sec:3+1daHydroQP}

In this section, we highlight the main ingredients and assumptions used in the 3+1d aHydroQP code used herein. First, we used smooth Glauber initial conditions, neglecting the effects of fluctuations.  We also used momentum-space isotropic initial conditions $\alpha_i(\tau_0)=1$ and assumed that $\eta/s={\rm const}$~\cite{Almaalol:2018gjh}.   For modeling the primordial hadron production and subsequent hadronic decays, we use a customized version of THERMINATOR 2 which allows for momentum-space anisotropies at freeze-out \cite{Chojnacki:2011hb}.  To determine the free parameters ($\eta/s$ and $T_0$), we  fit the pion, kaon, and proton spectra in the 0-5\% and 30-40\% centrality classes. Once the parameters were determined~\cite{Almaalol:2018gjh}, they can be used to compute other observables such as the elliptic flow and the HBT radii.  We note that our code is publicly available and can be obtained using Ref.~\cite{MikeCodeDB}.
 
\section{Phenomenological comparisons}
\label{sec:results}

In this section, we present  comparisons of the HBT radii predicted by aHydroQP with $\sqrt{s_{NN}}$ = 200 GeV Au-Au collision data. In this work, we continue our previous work \cite{Almaalol:2018gjh} and use exactly the same model parameters determined therein. It is worth mentioning that the parameters obtained from \cite{Almaalol:2018gjh} are $\tau_0 = 0.25$ fm/c, $T_0 = 455$ MeV, $\eta/s = 0.179$, and \mbox{$T_{\rm FO} = 130$ MeV}. For more details about the aHydroQP model and comparisons to RHIC hadron spectra, $v_2$, etc., we refer the reader to Ref.~\cite{Almaalol:2018gjh}.

The correlation function used to obtain the HBT radii is parametrized in terms of three Gaussians (the 3-D Bertsch-Pratt parameterization) \cite{Pratt:1986cc,Bertsch:1989vn}
\be
C(q,k)=  1+\lambda \, {\rm exp}[-R^{2}_{{\rm out}}q^{2}_{{\rm out}}-R^{2}_{{\rm side}}q^{2}_{{\rm side}}-R^{2}_{{\rm long}}q^{2}_{{\rm long}}] \, .
\label{eq:cform}
\ee
where $\lambda$ is the incoherence parameter. The longitudinal component ($R_{\rm long} $) is parallel to the beam axis (the z-axis), the out component ($R_{\rm out} $) is chosen parallel to the mean transverse momentum of  the pair  (the x-axis), and  the side component ($R_{\rm side} $) is perpendicular to both $R_{\rm long} $ and $R_{\rm out} $ (the y-axis). In a similar way, the relative momentum ($\vec{q}$=$\vec{p}_1$-$\vec{p}_2$) is decomposed into three components ($q_{\rm long} $, $q_{\rm out} $, $q_{\rm side} $) \cite{Florkowski:2010zz,Chaudhuri:2012yt}.  We have ignored final state interactions such as the Coulomb repulsion between the similar charged pion pairs.

\begin{figure}[t!]
\centerline{
\hspace{-1.5mm}
\includegraphics[width=1\linewidth]{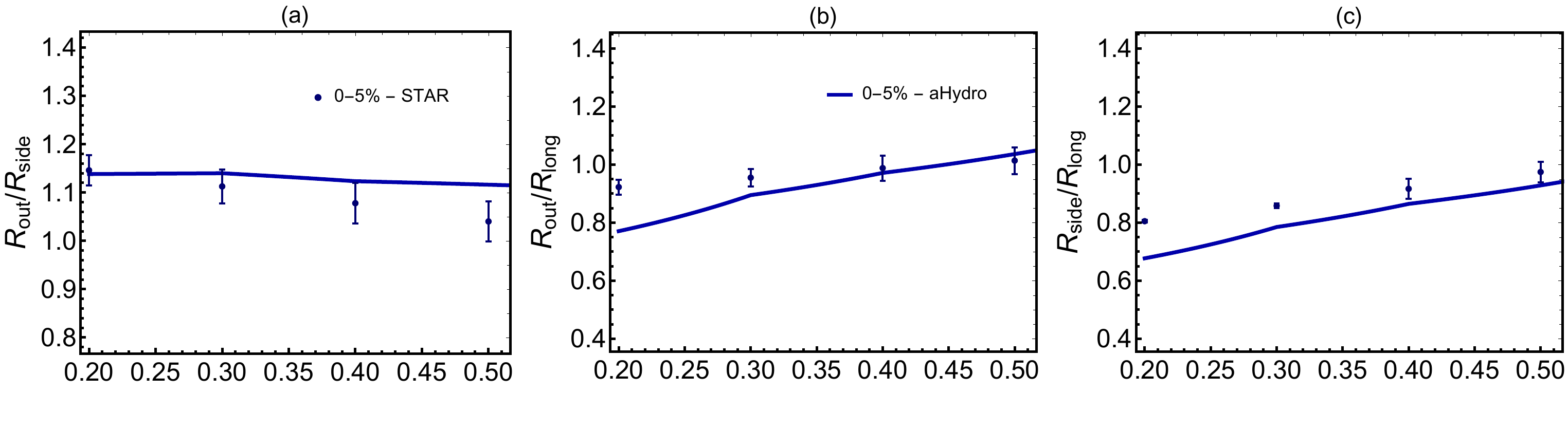}}
\centerline{
\hspace{-1.5mm}
\includegraphics[width=1\linewidth]{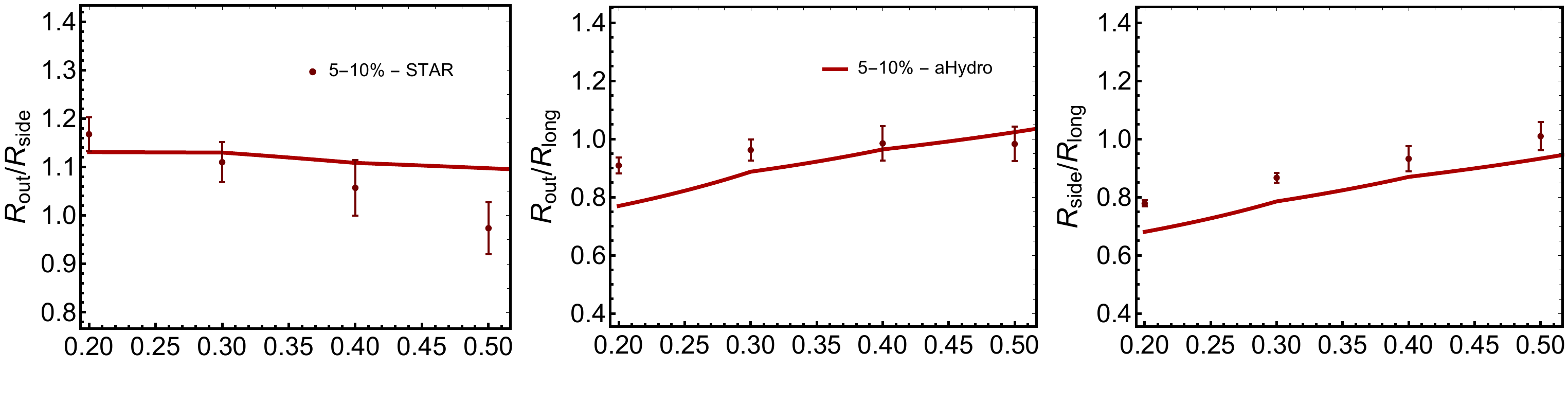}}
\centerline{
\hspace{-1.5mm}
\includegraphics[width=1\linewidth]{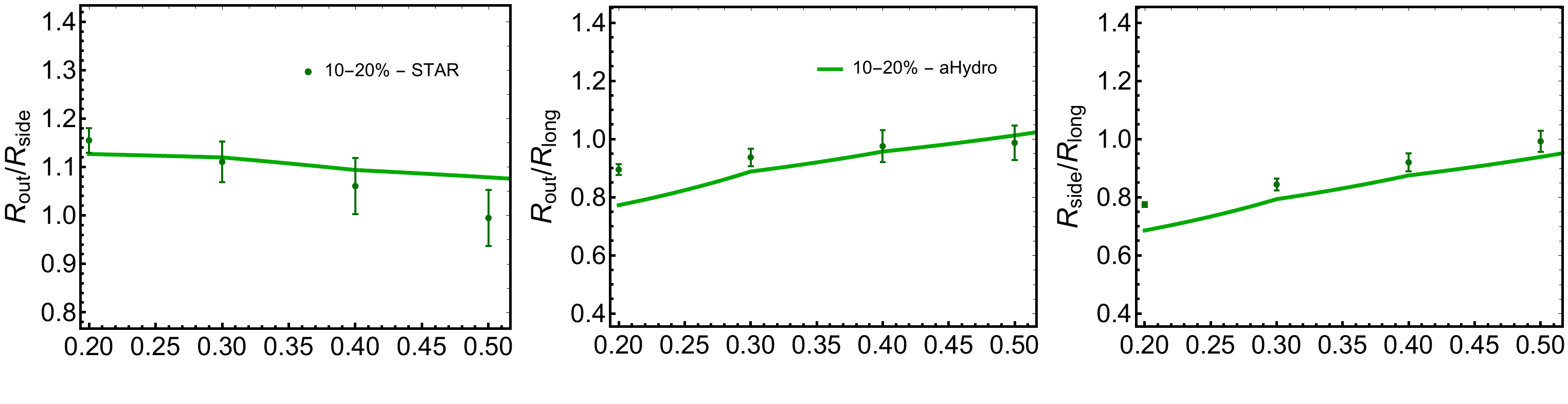}}
\centerline{
\hspace{-1.5mm}
\includegraphics[width=1\linewidth]{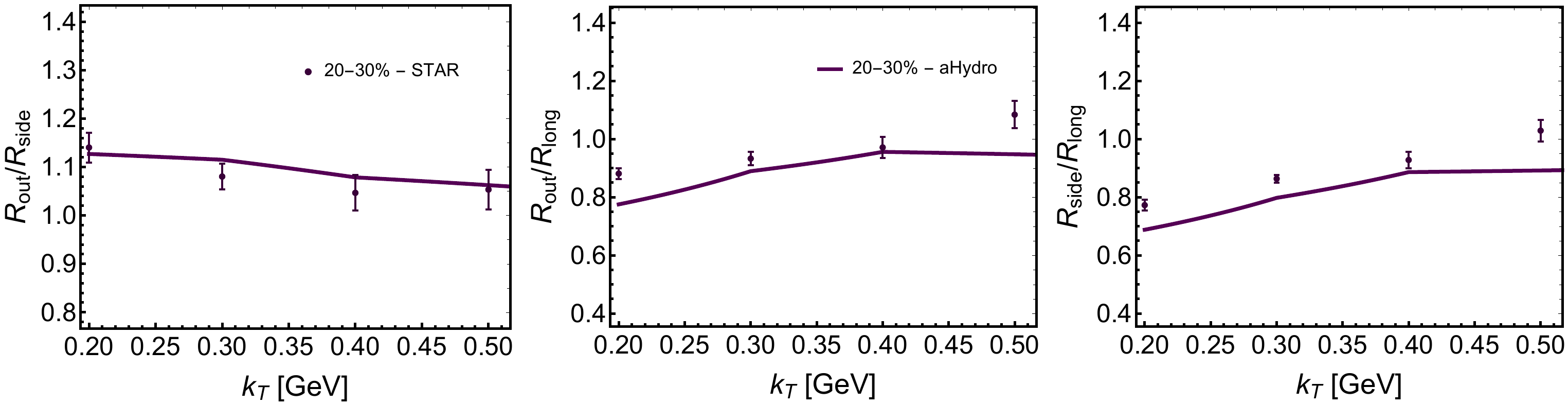}}
\caption{Ratios of HBT radii as a function of $k_T$ for $\pi^+ \pi^+ $ in the same centrality classes shown in Fig~\ref{fig:HBTSTAR}. The left, middle, and right columns show $R_{\rm out}/R_{\rm side} $, $ R_{\rm out}/R_{\rm long} $, and $R_{\rm side}/R_{\rm long}$, respectively. All results are for 200 GeV Au-Au collisions and the data shown are from the STAR collaboration \cite{Adams:2004yc}. }
\label{fig:HBTratiosSTAR}
\end{figure}

\begin{figure}[t!]
\centerline{
\hspace{-1.5mm}
\includegraphics[width=1\linewidth]{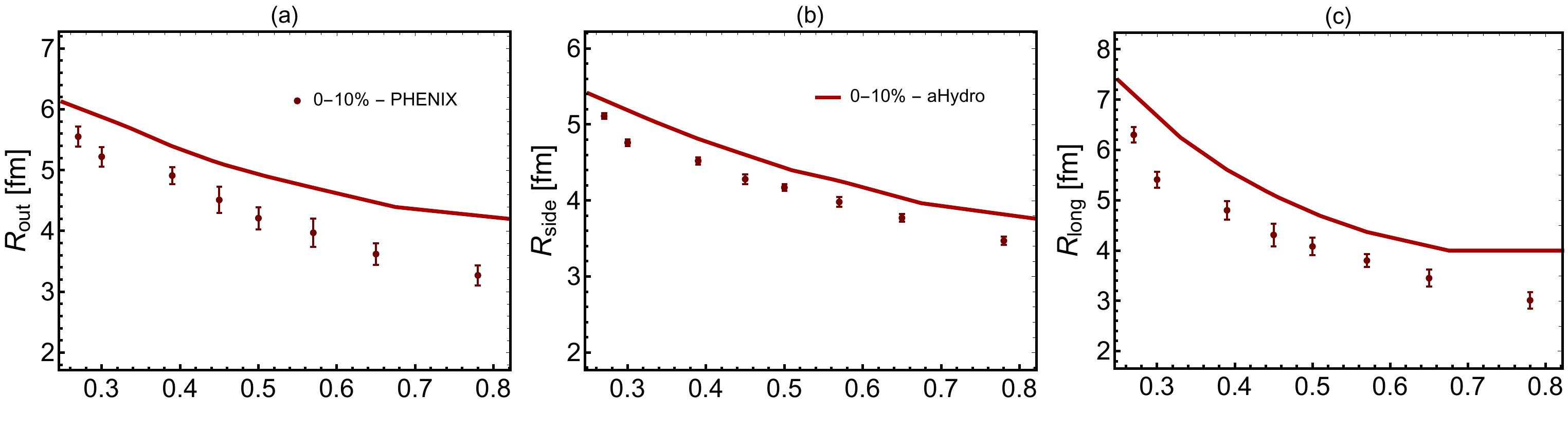}}
\centerline{
\hspace{-1.5mm}
\includegraphics[width=1\linewidth]{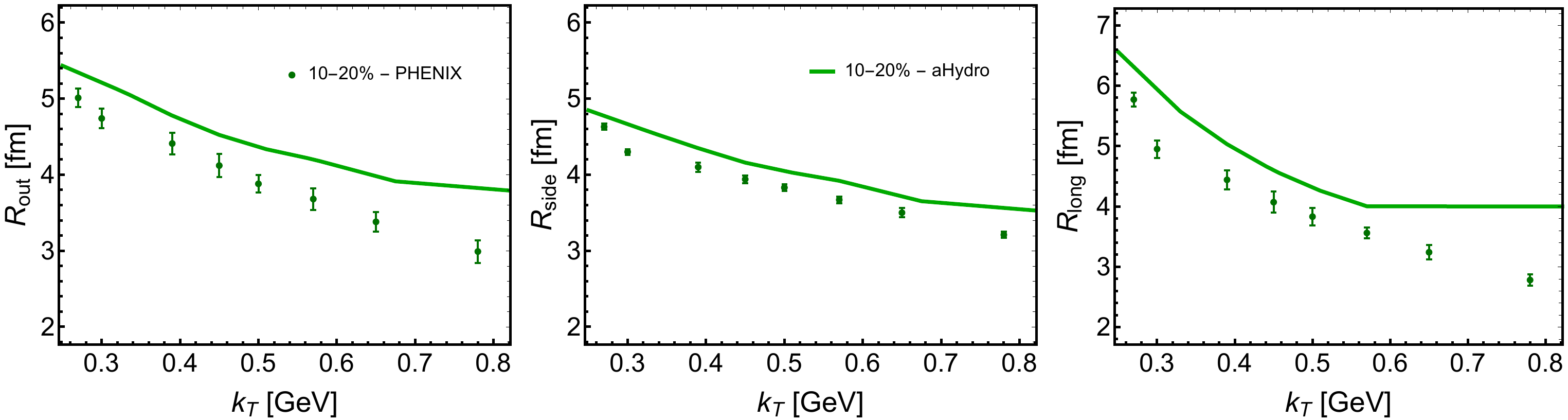}}
\caption{HBT radii as a function of  $k_T$ for $\pi^+ \pi^+ $ in the  0-10$\%$ and 10-20$\%$ centrality classes. The left, middle, and right columns show $R_{\rm out} $, $ R_{\rm side} $, and $R_{\rm long}$, respectively. All results are for 200 GeV Au-Au collisions and the data shown are from the  PHENIX collaboration \cite{Adare:2015bcj}. The $k_T$  cuts used were $0.2 < k_T < 2.0$~GeV~\cite{Adler:2004rq}.}
\label{fig:HBTPH}
\end{figure}

\begin{figure}[t!]
\centerline{
\hspace{-1.5mm}
\includegraphics[width=1\linewidth]{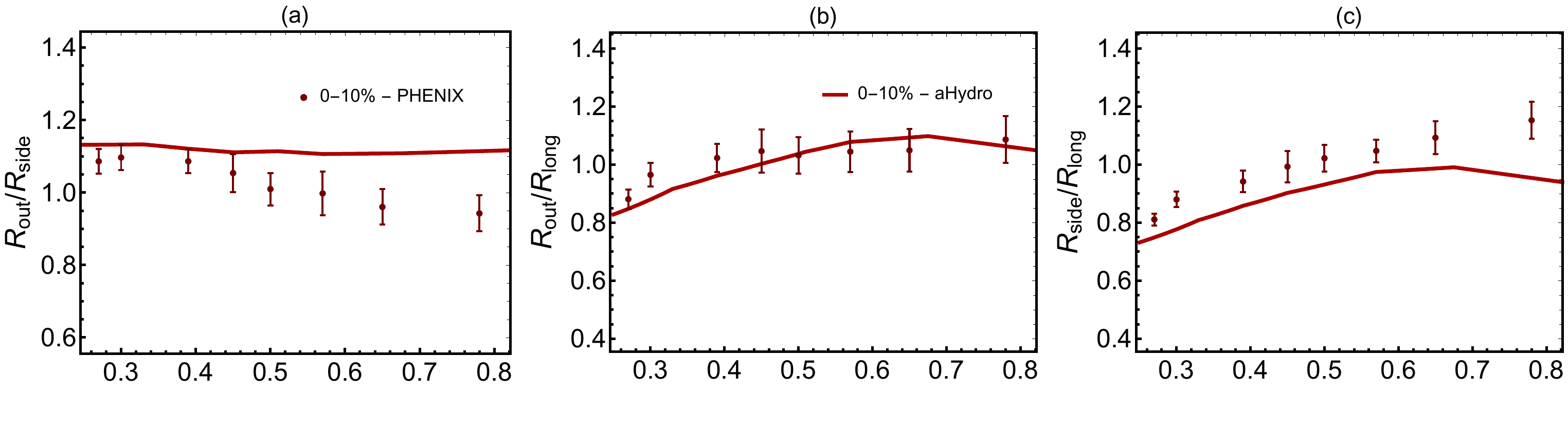}}
\centerline{
\hspace{-1.5mm}
\includegraphics[width=1\linewidth]{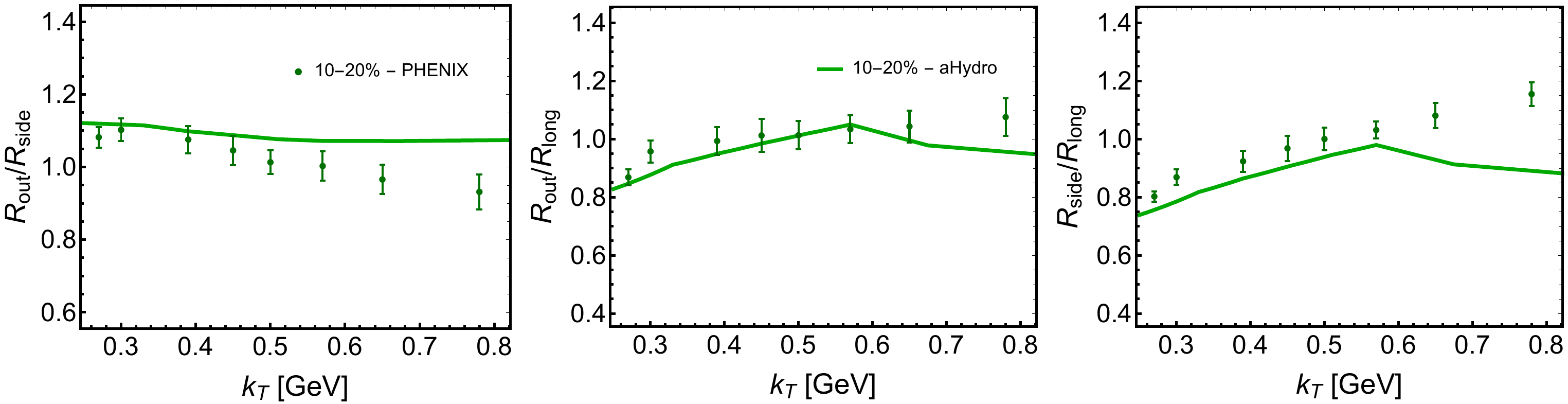}}
\caption{Ratios of HBT radii as a function of $k_T$ for $\pi^+ \pi^+ $ in the  0-10$\%$ and 10-20$\%$  centrality classes. The left, middle, and right columns show $R_{\rm out}/R_{\rm side} $, $ R_{\rm out}/R_{\rm long} $, and $R_{\rm side}/R_{\rm long}$, respectively. All results are for 200 GeV Au-Au collisions and data shown are from the PHENIX collaboration~\cite{Adare:2015bcj}.  }
\label{fig:HBTratiosPH}
\end{figure}

Now, let us turn to the comparisons between our model and the experimental results. In Figs.~\ref{fig:HBTSTAR}-\ref{fig:HBTratiosPH}, we show the $k_T $ dependence of the HBT radii and their ratios. As can be seen in Figs.~\ref{fig:HBTSTAR} and~\ref{fig:HBTPH}, the HBT radii tend to decrease with increasing $k_T $ due to the change in the size of the homogeneity regions.  In Figs.~\ref{fig:HBTSTAR}-\ref{fig:HBTratiosSTAR}, we show comparisons with STAR experimental results in four $k_T$ bins~\cite{Adams:2004yc}.  In Fig.~\ref{fig:HBTSTAR}, we show the predicted HBT radii by aHydroQP (solid lines) and the experimental results in four different centrality classes, 0-5\%, 5-10\%, 10-20\%, and 20-30\%. In the left, middle, and right columns we show $R_{\rm out} $, $ R_{\rm side} $, and $R_{\rm long} $ as a function of the mean transverse momentum of the pair $\pi^+ \pi^+$. In the left column, we see that our model overestimates the experimental data, particularly at $k_T \sim$ 0.5 GeV.  This is most likely due to the smooth initial conditions used herein.  Turning next to the middle column, we see that the agreement between our model and the STAR data is reasonable, however, quantitative differences remain.   Lastly, in the right column, which shows $R_{\rm long}$, we see that the agreement is clearly not as good as the other two columns especially at low $k_T$.

Next, in Figs.~\ref{fig:HBTratiosSTAR} we compare the ratios of the HBT radii. In this set of figures, in the left, middle, and right columns we show $R_{\rm out}/R_{\rm side} $, $ R_{\rm out}/R_{\rm long}  $, and $R_{\rm side}/R_{\rm long} $, respectively, as a function of $k_T$. As can be seen from this Figure, we find quite good agreement between our model and the experimental results for HBT radii ratios.  This is true for all ratios shown and in all centrality classes. We note that our model slightly underestimates the ratio $R_{\rm side}/R_{\rm long} $ due to overestimating $R_{\rm long} $ in Fig.~\ref{fig:HBTSTAR}.   The agreement with data found herein is similar to what was found in previous works using dissipative hydrodynamics, however, in this work we present comparisons in a larger range of centrality classes and compare with both STAR and PHENIX data~\cite{Broniowski:2008vp,Kisiel:2008ws,Pratt:2008sz,Pratt:2008qv,Bozek:2010er,Karpenko:2012yf,Bozek:2014hwa}.

Next, we show comparisons with PHENIX experimental results obtained from~\cite{Adare:2015bcj} where more $k_T$ bins are presented.  In Fig.~\ref{fig:HBTPH}, we show the comparisons of the HBT radii predicted by aHydroQP and the experimental results in  0-10\% and 10-20\% centrality classes. As in Fig.~\ref{fig:HBTSTAR},  the left, middle, and right columns show $R_{\rm out} $, $ R_{\rm side} $, and $R_{\rm long} $ as a function of the mean transverse momentum of the pair $\pi^+ \pi^+$. In a  similar way to what we have seen in the comparisons with STAR results, we see that our model overestimates the data especially at large $k_T$. As an example, in column (a), we see that there is an approximately 30\% difference between our model prediction and the experimental results at large $k_T \sim$ 0.8 GeV, while at at low $k_T \sim$ 0.3 GeV there is an approximately 20\% difference. Similar discrepancies  can also be seen in column (c) for the $ R_{\rm long} $ and in the bottom row for $ R_{\rm out} $ and $ R_{\rm long} $. However, as can be seen from column (b), in both centrality classes, we see that better agreement between our model prediction and the experimental results for $ R_{\rm side}$.

In Fig.~\ref{fig:HBTratiosPH}, we show the HBT radii ratios reported by PHENIX as a function of the pair mean transverse  momentum for $\pi^+ \pi^+ $ in the  0-10$\%$ and 10-20$\%$  centrality classes. As can be seen from these panels, the agreement is quite good up to $k_T \sim$ 0.5-0.6 GeV.   In the left column, which shows $R_{\rm out}/R_{\rm side} $, one sees that after $k_T \sim$ 0.5 GeV, our model predictions do not decrease enough to fully explain the experimental results. In the middle column,  which shows $ R_{\rm out}/R_{\rm long} $,  one sees that the agreement in the 0-10$\%$ centrality class is quite good, while in the 10-20$\%$ centrality class one sees that our model underestimates the data at large $k_T$. Finally, in the right column, which shows ($R_{\rm side}/R_{\rm long}$), one can see the model underestimates the data at large $k_T$.  Despite these quantitative differences, we note that overall the model does a good job in reproducing the systematic trends seen in the data.

From our comparisons with both STAR and PHENIX data, we see that our model overestimates $R_{\rm long}$ by on the order of 10-20\%.  This means that the longitudinal expansion of the QGP in aHydro is stronger than what is seen experimentally.  This is most likely due to the fact that aHydro was taken to have a strong initial boost-invariant Bjorken flow.  It would be interesting to study the impact of having a slightly weaker initial longitudinal flow profile.  We note here that the agreement between model and data presented here is quite similar to our prior comparisons between aHydroQP and ALICE results for 2.76 TeV Pb-Pb collisions~\cite{Alqahtani:2017tnq}, however, in the case of the PHENIX data, in particular, the experimental error bars are quite small which presents a challenge for the model.

\section{Conclusions and Outlook}
\label{sec:conclusions}

In previous works we have presented comparisons of 3+1d aHydroQP and relativistic heavy-ion collision experimental data obtained at different collision energies.  These past studies allowed us to fix a set of model parameters such as initial central temperature, shear viscosity, etc. for use in predicting additional observables.   In this paper, using those previously determined parameters, we presented predictions of the aHydroQP model for the femtoscopic HBT radii and their ratios at RHIC energies.   We showed comparisons of model predictions with data from both the STAR and PHENIX collaborations obtained using charged pion correlations.   We found reasonable agreement overall with experimental results in several centrality classes and as a function of the pair mean transverse momentum, in particular with respect to trends seen in the data.  There were, however, quantitative differences which were evident, for example, in $R_{\rm long}$.  These quantitative differences deserve further study.  In particular, it would be interesting to see how sensitive this observable is to the assumed initial longitudinal flow of the QGP.  Herein, for continuity with prior work, we assumed the initial longitudinal flow profile to be Bjorken flow, which is extremal.  Perhaps a modest reduction in the initial longitudinal flow would improve agreement with the experimental data beyond what has already been achieved.  Additionally, we have used smooth (non-fluctuating) Glauber initial conditions which is definitely too simplistic.  By including fluctuating initial conditions, one can expect to see reductions in the size regions of homogeneity, which may help to bring model predictions into better agreement with experimental data.  Looking to the future, we are working on comparisons with experimental data from 5.02 TeV Pb-Pb collisions, including HBT radii.  We also are working on including off-diagonal terms in the anisotropy tensor~\cite{Nopoush:2019vqc}, using more realistic (fluctuating) initial conditions, and using collisional kernels that go beyond the relaxation-time approximation~\cite{Almaalol:2018ynx,Almaalol:2018jmz}. 

\acknowledgments

M.\,A.\  is supported by the Deanship of Scientific Research at the Imam Abdulrahman Bin Faisal University under grant number 2020-080-CED. M. Strickland was supported by the U.S. Department of Energy, Office of Science, Office of Nuclear Physics under Award No. DE-SC0013470. This research utilized Imam Abdulrahman Bin Faisal (IAU)'s Bridge HPC facility, supported by IAU Scientific and High Performance Computing Center~\cite{Bridge}.
  
\bibliography{HBT_RHIC}

\end{document}